\pdfoutput=1
\documentclass{article}

\usepackage{arxiv}

\usepackage[ruled,linesnumbered]{algorithm2e}
\usepackage{booktabs}
\usepackage{amsmath}
\usepackage{multirow}
\usepackage{pifont}
\usepackage{xspace}
\usepackage{xcolor}
\usepackage{caption}
\usepackage{subcaption}
\usepackage{diagbox}
\usepackage{slashbox}
\usepackage{listings}
\usepackage[colorinlistoftodos]{todonotes}
\usepackage{enumitem}
\usepackage{fmtcount}
\usepackage{url}

\title{Building Reliable Cloud Services Using $\psharp$ (Experience Report)}

\author{
  Pantazis Deligiannis \\ 
  Microsoft Research, USA \\
  \texttt{pdeligia@microsoft.com} \\
\And
  Narayanan Ganapathy \\
  Facebook, USA\\
  \texttt{narg@fb.com} \\
\And
  Akash Lal \\
  Microsoft Research, India\\
  \texttt{akashl@microsoft.com} \\
\And
  Shaz Qadeer \\
  Facebook, USA\\
  \texttt{shaz@fb.com} \\
}

\newcommand{\psharp}{\textsc{P\#}\xspace}
\newcommand{\csharp}{\textsc{C\#}\xspace}

\newcommand{\figref}[1]{Figure~\ref{Fi:#1}}
\newcommand{\sectref}[1]{\S\ref{Se:#1}}

\newcommand{\Microsoft}{\textsc{Microsoft}\xspace}
\newcommand{\Azure}{\textsc{Azure}\xspace}
\newcommand{\AzureBatch}{\textsc{Azure Batch}\xspace}
\newcommand{\ABS}{\textsc{ABS}\xspace}
\newcommand{\AzureStorage}{\textsc{Azure Storage}\xspace}
\newcommand{\AzureSubscriptions}{\textsc{Azure Subscriptions}\xspace}
\newcommand{\AzureVMSS}{\textsc{Virtual Machine Scale Sets}\xspace}
\newcommand{\VMSS}{\textsc{VMSS}\xspace}
\newcommand{\AzureServiceBus}{\textsc{Azure ServiceBus}\xspace}
\newcommand{\AzureCosmosDB}{\textsc{Azure Cosmos DB}\xspace}

\newcommand{\Omit}[1]{}

\begin{document}
\maketitle

% ABSTRACT =================================
Cloud services must typically be distributed across a large number of machines
in order to make use of multiple compute and storage resources. This opens the
programmer to several sources of complexity such as concurrency, 
order of message delivery, lossy network, timeouts and failures,
all of which impose a high cognitive burden.
This paper presents evidence that technology inspired by formal-methods,
delivered as part of a programming framework,
can help address these challenges.
In particular, we describe the experience of several engineering teams in
\Microsoft \Azure that used the open-source $\psharp$ programming framework to
build multiple reliable cloud services.
$\psharp$ imposes a principled design pattern that allows writing formal
specifications alongside production code that can be systematically tested,
without deviating from routine engineering practices.
Engineering teams that have been using $\psharp$ have reported dramatically
increased productivity (in time taken to push new features to production)
as well as services that have been running live for months without
any issues in features developed and tested with $\psharp$.
%===========================================

\section{Introduction}
\label{Se:Introduction}

The public cloud offers access to an easily-managed, pay-on-use model of renting
compute and storage resources. Increasingly, many companies are moving their
business workloads to the cloud~\cite{forbes2018,forbes2019}.
This requires designing software services that execute on the cloud,
making effective use of the available resources.
However, developing such services is challenging as the cloud programming
environment is that of a traditional distributed system: service components
are spread across multiple virtual machines and data centers,
and communication must happen over the network.
To build a reliable cloud service, developers must defend against
all common pitfalls of distributed systems:
the concurrency from multiple executing processes, unreliable networks
(e.g., out-of-order delivery, or message loss/duplication),
as well as hardware/software failures. In this paper,
we refer to these combined challenges as sources of \textit{non-determinism}.
It is no surprise that the presence of such non-determinism leads to bugs
in production, causing tangible loss of business and customer
trust~\cite{amazon2012aws,google2014outage,tassey2002economic}.

The research community has made several attempts at finding distributed-systems bugs,
commonly through the use of \textit{systematic testing} tools. Examples include
\textsc{Chess}~\cite{musuvathi2008fair,musuvathi2008finding},
\textsc{MoDist}~\cite{yang2009modist}, \textsc{dBug}~\cite{simsa2011dbug} and
\textsc{SAMC}~\cite{leesatapornwongsa2014samc}.
These tools take over the non-determinism in a test environment
and control it to explore many different program executions. Both exhaustive
(up to a bound) and random explorations have proven to be effective.
In fact, folklore suggests that any distributed (or concurrent) system
when ``shaken'' carefully by a systematic testing tool would surely produce
bugs. However, despite of this success, there has been no visible change in
software development practices followed in the industry. Chances are that the next time
around a new system is built, it will be built in the same manner as before,
leading to the same kinds of bugs seen in previous systems. Without a change in
the software development process, the likely impact of any bug-finding technique
will be limited.

This paper presents evidence that the situation is not as grim as outlined above.
Tools and techniques suggested by the research community can indeed have
considerable impact in the industry for developing distributed systems.
We share experience from the adoption of the open-source $\psharp$ programming
framework~\cite{DBLP:conf/pldi/DeligiannisDKLT15,psharpgithub} in \Microsoft \Azure for building
production cloud services. $\psharp$ imposes a principled design pattern,
inspired from the actor-style of programming~\cite{Agha:1986:AMC:7929},
allowing the implementation to closely resemble its high-level design.
$\psharp$ also provides mechanisms for programmatically expressing
non-determinism and writing detailed safety and liveness specifications.
Finally, $\psharp$ comes with automated testing capabilities that
encapsulate the state-of-the-art in systematic testing. This enables high-coverage testing
of the production code against its specifications and provides deterministic
reproducibility of bugs. Putting these pieces together, $\psharp$ allows developers
to effectively iterate through the \textit{design-implement-test} cycle
faster than otherwise, leading to accelerated development.

To illustrate the benefits of using $\psharp$, we provide a detailed description
of \textit{PoolManager}, one of three core components of the
\AzureBatch service (\ABS)~\cite{azure-batch}
that was written from scratch in $\psharp$.
\ABS is a popular job scheduling and compute management service,
offered by \Azure, managing over \textit{hundreds of
thousands} of VMs on the cloud.
\ABS allows a user to create a collection of compute nodes and
schedule a parallel job across these nodes. PoolManager is
responsible for creating and managing the collection of compute nodes
(also called a \textit{pool}).
A previous version of PoolManager, developed over several years using standard
engineering practices, had an outdated design that was unable to manage the
increasing demands of \AzureBatch. It was hard to maintain and test, making feature
addition unacceptably slow. This prompted the \ABS team to
rewrite the PoolManager, this time adopting $\psharp$.

The \ABS engineers (both junior and senior) were able to move faster and be more
confident in their code changes because they could achieve high-coverage testing 
with $\psharp$.
Writing detailed specifications alongside production code became an integral
part of their daily development process.
The team reported that the coverage obtained with systematic testing
was much higher than even with days of stress testing.
$\psharp$ found \textit{hundreds} of bugs that
were fixed fast, and often without ever getting checked-in.
For a few bugs that we were able to snapshot, it was
unlikely that they would have been found using stress testing, or other
conventional testing methods, because they required several failures and timeout
events to interleave.

The \ABS team gained considerable confidence in $\psharp$ testing as the PoolManager
development proceeded: once a feature was tested with $\psharp$, it would
\textit{just work} when put into production. The current state of practice in the
team is that each code check-in to the \textit{master} branch must clear all
available $\psharp$ tests.
It was a unique experience for the team to get exhaustive
(in reality, high-coverage) testing of their code changes readily available
on their desktop as they were writing and integrating the code.
The debugging process was also significantly improved: each time $\psharp$ found a bug,
engineers could deterministically replay the buggy trace, attach a debugger,
set breakpoints and step-into the code.
The majority of the new PoolManager development took \textit{only six months},
considerably faster than the previous version. For some time, both
versions of PoolManager existed simultaneously.
During this time, the team had to add a new feature
(for supporting low priority preemptible VMs in the pools) to both versions.
The addition in the old PoolManager took six person months, whereas the addition
in the new $\psharp$ version took \textit{just one person month}.
The new PoolManager has now been operating for over a year with
\textit{no reported bugs in production} for $\psharp$ tested features.
There were occasional bugs, but they all pointed to features outside the
scope of their $\psharp$ tests.

The \AzureBatch PoolManager is the first production-scale system, to the best of our
knowledge, to have been developed along with continuous
validation of safety and liveness specifications.
The experience of the \ABS team was not isolated. Given their success,
several other teams in \Azure adopted $\psharp$ in their engineering process.
Currently, $\psharp$ has been used in \Azure for building nine production services,
with several more services in the planning stage (\sectref{Experience}).
Furthermore, $\psharp$ has $100\%$ user retention so far: once a team started
using it, they have continued to use it for writing new cloud services.

The main contributions of this paper are as follows:
\begin{itemize}
\item We present PoolManager, the first production-scale system to have been
developed simultaneously with continuous validation of the actual code against
its safety and liveness specifications; both design and implementation was done
by engineers, not researchers.

\item PoolManager is a stateful microservice that requires storing its state reliably
so that it can be restored after a failure (known as a \textit{failover}). Getting the
failover logic correct is often hard. We give a novel methodology for failover
testing (\sectref{PoolManagerTesting}).

\item We describe several improvements to $\psharp$ that were necessary to support
  industry-scale usage (\sectref{ImprovementsToPSharp}).

\item We discuss the experience of several \Azure engineering teams with using $\psharp$,
  for building highly-reliable cloud services (\sectref{Experience}).

\end{itemize}

\noindent
The rest of this paper is organized as follows:
\sectref{Overview} provides background;
\sectref{PoolManagerDesign} outlines the design and implementation
of the PoolManager using $\psharp$;
\sectref{PoolManagerTesting} focuses on testing of the PoolManager;
\sectref{ImprovementsToPSharp} lists the improvements made to $\psharp$
to support the development of production systems;
\sectref{Experience} summarizes the experience of several \Azure engineering
teams with using $\psharp$; and finally
\sectref{RelatedWork} presents related work.

\section{Overview}
\label{Se:Overview}

\subsection{The \AzureBatch Service (\ABS)}
\label{Se:ABS}

\ABS is a popular generic job scheduling service offered by \Microsoft \Azure~\cite{azure-batch}.
\ABS allows a user to execute a parallel job in the cloud. The job can
consist of multiple tasks with a given set of dependencies. \ABS will execute the
tasks in dependency order while attempting to exploit as much parallelism
between independent tasks as possible.
Unlike distributed schedulers such as Apache \textsc{Yarn}
\cite{conf/cloud/VavilapalliMDAKEGLSSSCORRB13} and \textsc{Mesos}
\cite{Hindman:2011:MPF:1972457.1972488} that typically require to be installed
on a pre-created set of VMs, \ABS integrates scheduling with VM
management. \ABS can \textit{auto-scale} (i.e., spin up and down) the number of
created VMs based on the needs of each job as well as a variety of parameters such as
CPU, memory and I/O metrics on VMs, and preemption rate.

\begin{figure*}
\centering
\includegraphics[width=0.9\linewidth]{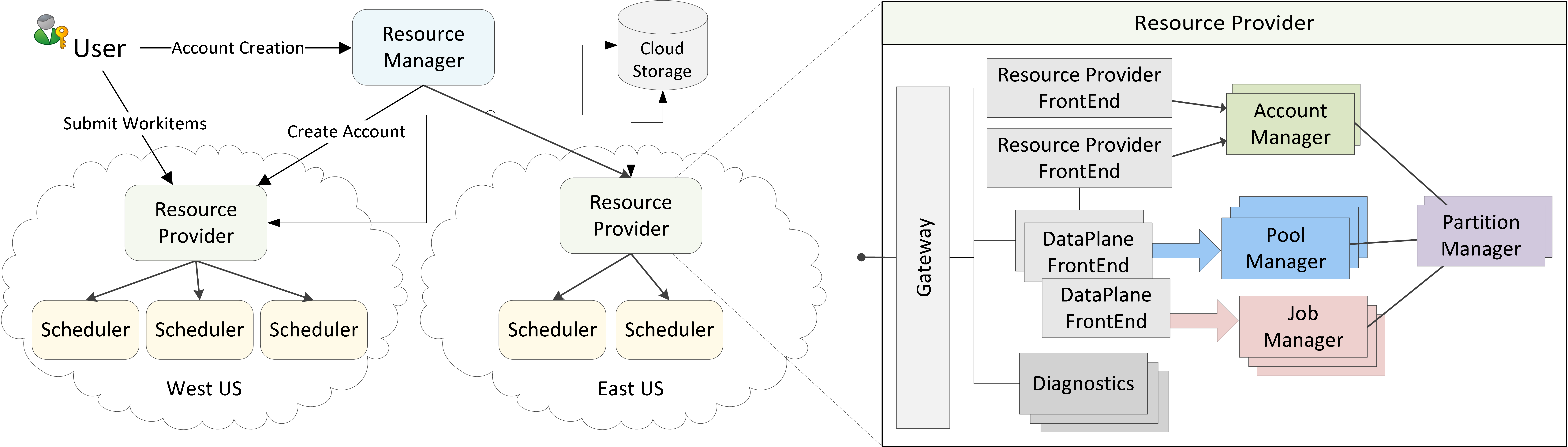}
\vspace{-1mm}
\caption{The \AzureBatch service architecture on the left,
and the Resource Provider architecture on the right.}
\label{Fi:Architecture}
\vspace{-1mm}
\end{figure*}

The high-level architecture of \ABS is shown on the left side of \figref{Architecture}.
Each \textit{region} (i.e., a geographical location that hosts one or more data
centers) has a resource provider and several \textit{schedulers}.
A breakdown of the resource provider is shown on the right side of \figref{Architecture}.
The resource provider has a front-end or gateway service that routes requests to
back-end managers that support the CRUD operations for specific entities:
\textit{user accounts}, \textit{pools} and \textit{jobs}, with relevant data stored in
\AzureStorage~\cite{azure-storage} for persistence.

\ABS is a multi-tenant service and a user account is the multi-tenant isolation boundary.
Each account is associated with a \textit{quota} that limits the amount of compute
resources that can be allocated for a scheduled job.
All resources used by \ABS for executing a job are billed to the corresponding account.
After creating and registering an account, a user can create a \textit{pool} that
refers to a collection of compute nodes (VMs). The sum total of cores of all VMs
across all pools associated with an account must be less than the corresponding core
quota limit of the account. The pool can be of fixed size, or set to auto-scale.
Once a pool is created, the user can submit a job to schedule on the pool.
Each of the account, pool and job managers are multi-instance partitioned services
(partitioned by account). The partitions themselves are managed by the partition manager.
The design of the partition manager is out of scope for this paper but the various
services have to honor partition manager requests to start/stop/split/merge partitions.

\paragraph{PoolManager}
At the heart of the \ABS functionality is the microservice component called
PoolManager. The PoolManager interacts with many other components in the system,
such as the job manager, \AzureStorage~\cite{azure-storage} for storage,
\AzureVMSS (\VMSS)~\cite{AzureVMSS} for VM management and
\AzureSubscriptions~\cite{azure-subs} for billing accounts.
\ABS needs to respond to auto-scale requirements very rapidly. To achieve
this, it must support functionality to cancel outstanding operations so that
further changes to resources can be made. The
functionality must be provided with low latency, high throughput, high
availability and scale. Note that all services that \ABS interacts with are
publicly available and \ABS uses the same APIs that are published externally.
This implies that \ABS must obey all the rules and limitations enforced by these services.
This point was an important design consideration, especially for \ABS quota and
billing management.
% subscription management.

The PoolManager component had to be redesigned for a variety of reasons. The
old design split the work of creating pools between the PoolManager and the
scheduler. The PoolManager managed pool entities and quotas while the scheduler did the actual
allocation of the VMs for the pools. This design itself was an evolution of a previous
design where the scheduler cached a pool of VMs. In that scenario, allocating a pool
was not about creating new VMs but rather picking from a set of free VMs.
Caching of VMs was no longer feasible for \ABS: as its usage increased, customers wanted
VMs of different sizes and OS images, etc., so it became too costly to hold the VMs in the
scheduler. The old design made the scheduler very complicated.
It was also harder to dynamically scale up and down the number of schedulers because
they were involved in VM management. The goal of the redesign
was to move the quota management and the actual VM allocation to a single component
(PoolManager) where it can be partitioned and scaled independently of the scheduler component.
This helped the scheduler become a very lightweight component that
can be spun up and down quickly, as it now only focuses on job scheduling.
The redesign also allowed the \ABS team to easily
incorporate different types of scheduling policies.

The old PoolManager did have some unit and integration tests but the \ABS
team felt that the tests did not provide much
confidence in the overall reliability of the PoolManager.
It was important to remedy this situation as well. The PoolManager is a stateful
component that operates in a distributed-systems environment. Testing of such
components is challenging. For instance, the VM hosting a PoolManager instance
may fail or reboot without warning. Operations on pools
are long-running asynchronous activities, thus, the developer must anticipate and account for
failures that can happen in the middle of such an operation. We refer to the part of a
program's design that deals with recovery from failures as \textit{failover
logic} and testing for its correctness as \textit{failover testing}. Failovers
are not the only challenge: one must also correctly deal with issues such as message
re-orderings, timeouts and error handling (when interacting with other services).
Due to inadequate testing technology, most often, errors arising from these types of issues
are discovered late in the development cycle, or even after deployment when they
are very costly to debug and fix. Ideally, we should be able to
discover and fix these kinds of issues well before the software is deployed in
production.

A further requirement of the PoolManager design was that the entire code base
should be asynchronous and non-blocking. This is to ensure that PoolManager
remains responsive to cancellation requests: the user is allowed to cancel an
outstanding operation at any time. In the old design, there were dedicated threads
that blocked synchronously and when the process
ran out of threads, the system could not process more requests.

\subsection{An Introduction to the \psharp Framework}
\label{Se:PSharp}

$\psharp$~\cite{DBLP:conf/pldi/DeligiannisDKLT15,psharpgithub} is an open-source
actor-based~\cite{Agha:1986:AMC:7929} .NET programming framework.
An actor, which is the unit of concurrency, in $\psharp$ is called a
\texttt{machine}.
A $\psharp$ program can dynamically create any number of \texttt{machines}
that execute concurrently with each other and communicate via messages
called \texttt{events}.
Each \texttt{machine} is equipped with an inbox where incoming
\texttt{events} get enqueued, and executes an event-handling loop
that waits for \texttt{events} to arrive and processes them
sequentially one after the other.
A \texttt{machine} can internally define a \textit{state machine}
structure for programmatic convenience (which is where the term
\texttt{machine} comes from). However, this feature is merely
syntactic sugar and not central to the value provided by $\psharp$.

$\psharp$ has a higher-level concurrency model compared to using threads and locks.
A \texttt{machine} encapsulates its own state that is not shared with
other \texttt{machines}, and synchronization is limited to sending \texttt{events}.
This means that all communication points between \texttt{machines}
are clearly marked in code (as opposed to using shared memory where
communication happens implicitly each time shared data is accessed).

$\psharp$ is designed in a manner that allows robust testing of non-deterministic
systems. To this effect, $\psharp$ requires developers to explicitly
declare all non-determinism present in their code, after which they can use
the $\psharp$ tester to exercise (in the limit) all possible behaviors
of a given test case.
The tester understands the non-determinism that arises from concurrency
between \texttt{machines}. It uses hooks into the $\psharp$
runtime to control the scheduling of \texttt{machines}.
%(Building the same functionality for shared-memory programming models usually requires code instrumentation, which has its own
%tool-building challenges.)
There can be other forms of non-determinism in the code. $\psharp$ exposes
an API to generate unconstrained Boolean and integer values.
We refer to this API as \textsc{NonDet} in the rest of this paper.
It is the responsibility of the programmer to \textit{model} the non-determinism
in their code using this API. We illustrate this point using an example.

Consider building an application that requires running multiple
$\psharp$ \texttt{machines} distributed over a network.
The developer is interested in testing the implementation against a \textit{lossy}
network, as the code must work correctly even if the
network arbitrarily drops messages.
The developer first writes a test that initializes all $\psharp$ \texttt{machines}
under a mocked distributed environment so that the code can execute in a
single-process setting.
This mocked environment can model the network to express its lossy behavior.
\figref{LossyNetworkMock} shows an illustration for such a mock.
The application (not shown) is designed against an interface
of the network (\texttt{INetworkingService}), which is then
mocked (\texttt{MockNetworkingService}) for testing purposes. The mocked
method \texttt{SendMessage} calls \textsc{NonDet} to decide if it is going
to deliver the \texttt{event} or not. When it must deliver, it directly
addresses the target \texttt{machine} and delivers the \texttt{event} via
a $\psharp$ \texttt{Send}.
% (Similarly, it is easy to model other scenarios such as a network
% that can duplicate or reorder messages.)
Once external dependencies are substituted with mocks to make the test self-contained
(as in standard unit-testing), the $\psharp$ tester repeatedly executes the test,
each time exploring a different interleaving of concurrent actions as well as
resolving \textsc{NonDet} calls with different values.

\begin{figure}[t]
\begin{lstlisting}[language={[Sharp]C},basicstyle=\scriptsize]
interface INetworkingService {
  void SendMessage(string endPoint, Event message);
}

class MockNetworkingService : INetworkingService {
  // Map: Endpoint -> Machine hosted at the endpoint
  Dictionary<string, MachineId> machineMap;

  void SendMessage(string endPoint, Event message) {
    if(PSharpRuntime.NonDet()) {
      PSharpRuntime.Send(machineMap[endpoint], message);
    }
  }
}
\end{lstlisting}
% \vspace{-3mm}
\caption{A $\psharp$ mock for a lossy network.}
% \vspace{-2mm}
\label{Fi:LossyNetworkMock}
\end{figure}

The $\psharp$ tester supports many state-of-the-art search strategies inspired
from the systematic testing literature~\cite{DBLP:conf/asplos/BurckhardtKMN10,
DBLP:conf/fmcad/MudduluruDDLQ17,DBLP:conf/sigsoft/DesaiQS15},
and makes it easy to add new strategies as the research community
comes up with new algorithms.
By default, the tool is configured to execute a portfolio of search
strategies in parallel to provide the best coverage to the user.

In addition, $\psharp$ provides support for writing functional
\textit{specifications} of the code.
These specifications are written in the form of \texttt{monitors}.
A \texttt{monitor} can only observe the execution of a program but cannot influence it.
Syntactically, this means that it can receive messages from any \texttt{machine},
but cannot send messages. A \texttt{monitor} makes it easy to assert conditions
that span multiple \texttt{machines}. 

A \texttt{monitor} can also encode a \textit{liveness} property
to check if a system is making progress.
A \texttt{monitor} can indicate their \textit{temperature} as either
\textit{hot} or \textit{cold}.
Given such \texttt{monitor}, the $\psharp$ tester searches for an execution
where the \texttt{monitor} remains in a hot state for an ``infinite'' amount
of time (in reality using a heuristic) without transitioning to
a cold state~\cite{DBLP:conf/fmcad/MudduluruDDLQ17}.
For instance, consider a replication protocol that is required to maintain,
say, three replicas of some data.
The developer can write a \texttt{monitor} that turns \textit{hot} when
one replica fails (and the count falls below three)
and turns \textit{cold} when three replicas have come up.
This \texttt{monitor} encodes the specification that the protocol
\textit{eventually} creates the required number of replicas,
even in the presence of failures.
Such a specification was previously used for finding bugs in a storage system~\cite{DBLP:conf/fast/DeligiannisMTCD16}.
It is common for distributed systems to
have liveness requirements~\cite{lamport1994temporal}.

To summarize, a system developed using $\psharp$ typically implies three activities.
First, the system itself must be written using the $\psharp$ concurrency model.
The $\psharp$ runtime provides APIs to create \texttt{machines}
and send \texttt{events}.
Second, external dependencies must be mocked and
all sources of non-determinism must be expressed 
via \textsc{NonDet} calls.
Third, the user writes tests (exercising the system under a workload)
and specifications for asserting correctness.

\section{Implementing the PoolManager in \psharp}
\label{Se:PoolManagerDesign}

This section outlines the design of the \ABS PoolManager and
how it is implemented using $\psharp$.
The goal is only to provide enough details to impress the complexity of the service,
justifying the need to use $\psharp$, and not to give an exhaustive account of
the system. 
We believe the core reasons behind the complexity are common to many 
cloud systems.

PoolManager exposes APIs for creating a pool, resizing or deleting an
existing pool, as well as canceling a previous resizing operation.
We begin by explaining key external services that PoolManager relies on
for implementing its functionality before getting into the PoolManager
design.

\subsection{External Services}
\label{Se:ExternalServices}

\paragraph{\AzureStorage}
PoolManager operations are naturally long-running because
the creation or deletion of VMs takes time
(in the order of seconds to a few minutes).
If PoolManager fails while creating, say, a pool of size $10$
after it has already allocated $3$ VMs, then after restarting,
it must resume the pending operation and allocate only the
remaining $7$ VMs.
It is important to not lose track of previously allocated VMs
(i.e., they must be part of some pool), else \ABS risks allocating
resources that will never be subsequently used.

Anticipating failures in the middle of an operation, the PoolManager
records its progress using \AzureStorage~\cite{azure-storage},
a highly available and reliable cloud-scale storage system.
\AzureStorage offers a key-value storage interface. PoolManager uses REST APIs
to read and write information about pools, VMs, jobs, tasks, quota management,
etc., as entities (rows) in storage.
\AzureStorage also provides opportunistic concurrency control using
\textit{entity tags} or \textit{ETags}. These are metadata attached to each row.
A client can do a conditional write to a row: the row is updated only if the
user-provided ETag matches the current value in the row.

\paragraph{\AzureSubscriptions}
All \Azure resources that a customer allocates belong to
a subscription, which is a billing entity.
Subscriptions, which are managed by the \AzureSubscriptions service,
can contain accounts for services such as \AzureBatch and \AzureStorage.
\Azure imposes limits on how many operations a subscription
can perform on resources.
Similarly, there are limits on the number of cores one can allocate
via VMs in a single subscription.
The provided limits are too restricting for running \ABS workloads.
As \ABS is built on public \Azure services and needs to allocate resources,
it has to own a set of subscriptions (with fairly large limits)
and use them to manage resources.
By spreading resources across many internal subscriptions, \ABS scales
beyond what can be achieved with a single subscription.

\paragraph{\VMSS}
\ABS uses \VMSS~\cite{AzureVMSS} 
in order to allocate VMs for creating a pool. \VMSS offers a service for
allocating a collection of VMs, which \ABS further wraps into the concept of a
pool, for the following reasons:
\begin{enumerate}
\item \ABS ties scheduling with resource provisioning. If \ABS has a
pending request to shrink a pool, and a VM finishes running a task,
then \ABS proceeds to collect and free the VM. This tight
coupling between scheduling and resource provisioning is an important
value proposition of \ABS.

\item \VMSS imposes VM creation limits per subscription.
\ABS pools can be much larger than these limits.
\VMSS also limits the number of operations per subscription.
\ABS spreads out pool creations across many subscriptions
to speed up deployment.

\item \ABS supports pool operations such as stop-resize.
This is not supported by \VMSS.
When a user issues a stop-resize operation, \ABS moves the
corresponding \VMSS operations to the background
(the customer is not charged) and deletes the extra VMs allocated.
The stop operation offered by \ABS allows customers to respond
to compute demand more quickly.
\end{enumerate}

\noindent
The \textit{DeploymentManager} (which is also written in $\psharp$)
is a microservice component of \ABS that interfaces with \VMSS.
PoolManager uses the DeploymentManager service to create, grow, shrink and delete
individual \VMSS collections, also called \textit{deployments} in the rest of the paper.

\subsection{PoolManager Design}

\paragraph{Design requirements}
Central to the PoolManager is the need to manage quotas. Each user has a quota
on the maximum number of VMs they can allocate for running their jobs.
Further, \ABS internally manages multiple \Azure subscriptions, each one
tied to one region, which limits the number of VMs that can be allocated
in that region.
Thus, a VM can be allocated for a user only when the user-quota has not
been exceeded, and there is some subscription (in some region) whose
quota has not been exceeded.

In addition to pool operations such as create and resize, PoolManager
can recover VMs that are determined to be unhealthy.
The unhealthy signal comes from the \ABS scheduler and the PoolManager,
in response, deletes the unhealthy VM (by signalling to \VMSS)
and allocates a new one.

\begin{figure*}
\centering
\includegraphics[width=0.9\linewidth]{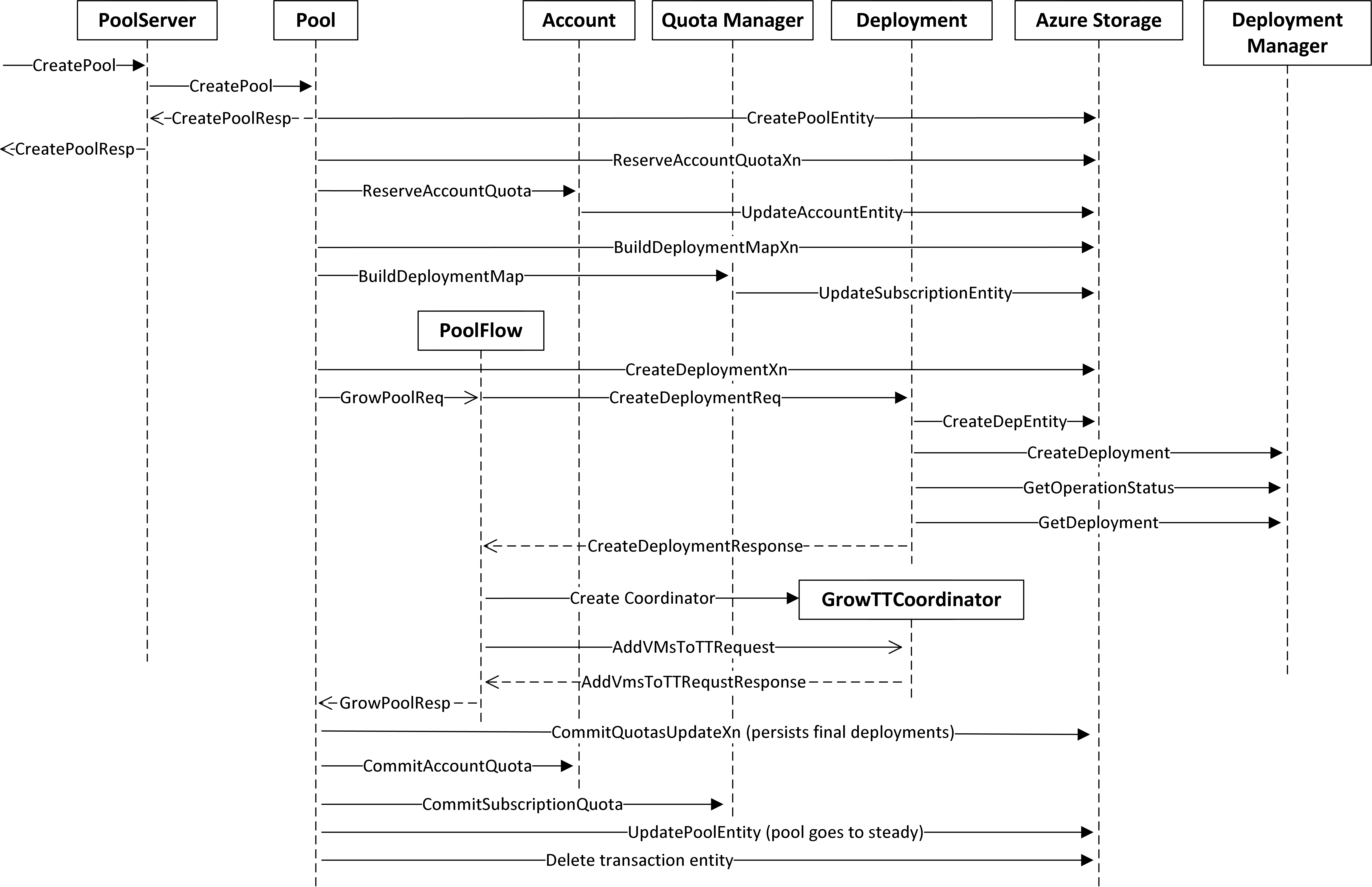}
% \vspace{-2mm}
\caption{The PoolManager workflow for creating a pool.}
% \vspace{-1.5mm}
\label{Fi:CreatePool}
\end{figure*}

\paragraph{$\psharp$ machines}
The PoolManager implementation has over $30$ types of $\psharp$ \texttt{machines},
totalling over $50$K lines of code. 
Each \texttt{machine} is designed to manage the lifetime of a particular resource
or to execute a workflow that implements a sub-operation.
For example,
the \texttt{Pool} \texttt{machine} manages a single pool,
the \texttt{PoolServer} \texttt{machine} manages a collection of
\texttt{Pool} \texttt{machines},
the \texttt{PoolFlow} \texttt{machine} allocates resources for a pool,
the \texttt{Deployment} \texttt{machine} manages a single deployment,
the \texttt{Account} \texttt{machine} manages a single account,
the \texttt{QuotaManager} \texttt{machine} manages quota
requirements and decides how to allocate across subscriptions.

\subsection{PoolManager Operations}
\label{Se:PoolManagerOperations}

\paragraph{CreatePool}
To create a pool, PoolManager goes through the following steps:
$(1)$ persists the pool properties and puts the pool in resizing state, 
$(2)$ checks if enough quota is available and tentatively reserves it,
$(3)$ allocates resources and creates VMs to match the required pool size,
$(4)$ persists deployment and VM information,
$(5)$ informs the scheduler about the created VMs so that it can start
scheduling job tasks on the VMs,
$(6)$ commits the revised leftover quota,
$(7)$ updates the pool properties to the final count of resources and
puts the pool in \textit{steady state}. An operation is deemed completed
once the pool reaches steady state.

\figref{CreatePool} shows the workflow that implements this
operation to highlight its complexity.
Each vertical line corresponds to a $\psharp$ \texttt{machine}.
Arrows represent exchange of \texttt{events} between \texttt{machines}.
The \texttt{AzureStorage} \texttt{machine} wraps calls to
the \AzureStorage service.
Arrows to \texttt{AzureStorage} represent a read or write of
persistent data.
The \texttt{DeploymentManager} \texttt{machine} wraps \VMSS,
and arrows to this \texttt{machine} represent \VMSS operations.

It is important to note that the workflow is a simplified linear view
of the PoolManager execution. The reality is even more complex because of
two reasons.
First, the workflow executes in parallel (all \texttt{machines} are
running concurrently), so their responses can arrive in different orders.
Second, error-handling code is pervasive.
Each operation, especially when interacting with external services,
can return an error code (or time out), which must be handled appropriately.
Due to these reasons, the asynchronous programming model of $\psharp$ fits
naturally with the PoolManager requirements: the \texttt{machines} send
out requests and field responses as they arrive asynchronously,
instead of blocking each time for a response.

\paragraph{ResizePool}
A resize-pool operation is similar to creating a pool, except that it 
may have to grow or shrink existing deployments, in addition to creating new
ones.
The resize-pool workflow goes through the following steps:
$(1)$ persists the new target values and puts the pool in resizing state,
$(2)$ checks quotas, if the resize involves growing the pool,
$(3)$ for grow operations: allocates resources and creates or grows
deployments, then persists the updated information, 
$(4)$ for shrink operations: works with scheduler instances to identify 
VMs that can be deleted,
$(5)$ commits the revised quota,
$(6)$ updates pool properties to reflect final counts and puts the
pool in steady state.

\paragraph{CancelResize}
The PoolManager allows only one resize or delete operation per pool
at a time. To stop this operation, the user can
issue a cancellation that goes through the following steps:
$(1)$ stops operations that were creating or resizing deployments,
$(2)$ updates the pool size to the previous size plus (minus) any deployments
whose creation (deletion) has already committed, and
$(3)$ puts the pool in steady state.

After a cancellation is carried out, there may be deployments with extra
VMs that have not been freed. These are termed \textit{rogue} VMs.
After the pool goes to steady state, a \textit{stabilize deployment} operation
starts in the background that asynchronously removes any rogue VMs.
This operation first identifies such deployments and puts them in
a \textit{stabilizing} state (subsequent resize operations
skip deployments that are in this state).
It then waits for pending operations on the deployment to complete
before issuing fresh operations to remove the rogue VMs.
The stabilize-deployment operation also needs to persist progress to storage;
on failover, the stabilization is resumed.

An additional requirement is to limit the total number of
rogue VMs across all deployments.
The \texttt{QuotaManager} \texttt{machine} implements this logic:
it aggregates the rogue-VM count across all its deployments.
If this number exceeds a threshold, the PoolManager stops
new cancellation requests until the rogue-VM count comes down.

\section{Testing the PoolManager with $\psharp$}
\label{Se:PoolManagerTesting}

As illustrated in \sectref{PoolManagerDesign}, PoolManager involves multiple
different operations that can be running concurrently at any point in time.
Furthermore, PoolManager has to deal with failures that can happen unexpectedly,
and has to correctly resume all pending operations after a failover.
Testing such a complex system is where using $\psharp$ truly
makes a difference.

\subsection{PoolManager Specifications}
\label{Se:PoolManagerSpec}

Any testing effort with $\psharp$ must start by writing a specification
that defines all valid system behaviors.
The PoolManager specification is around $1,700$ LoC and is
written as a $\psharp$ \texttt{monitor} (\sectref{PSharp})
that captures the following properties:

\paragraph {Liveness properties}
\begin{enumerate}
\item For a given pool, if the last client operation was a resize to size $N$, then
the pool \textit{eventually} reaches steady state with size $N$.
\item For a given pool, if the last client operation was a delete, then the pool is
\textit{eventually} removed and all its allocated VMs are returned back to
\VMSS.
\item For a given pool, all pending stabilization operations must eventually
complete.
\item For a given pool, all pending delete operations must eventually
complete.
\item For a given pool, all pending recovery operations must eventually
complete.
\end{enumerate}

\paragraph{Safety properties}
\begin{enumerate}
\item In steady state, the state of PoolManager is in sync with \VMSS.
In particular, if the PoolManager believes that a pool has VMs
$\{v_1, v_2, \cdots v_n\}$, then these VMs have indeed been allocated by
\VMSS to the PoolManager.
\item For every successful create-pool request, a pool entry
is created in \AzureStorage.
\item For every successful resize-pool request, the pool target matches
what is requested in the corresponding \AzureStorage entry.
\item For every successful delete-pool request, the pool entry
is deleted from \AzureStorage.
\item For every pool, every deployment enumerated in the \AzureStorage pool entry
is present in the \AzureStorage deployment entry.
\item Every deployment in the \AzureStorage deployment entry belongs to a pool.
\item Every VM in the \AzureStorage VM entry belongs to a pool.
\end{enumerate}

\noindent
Importantly, the above properties must hold even after a failover.
Checking the PoolManager against this specification, especially to get coverage
of corner-case behaviors, is a challenging task for several reasons.
For instance:

\begin{enumerate}
\item The PoolManager is a concurrent program; its various operations may
interleave in many different ways.
\item The PoolManager interacts with multiple external services and it must be able
to handle any valid response from those services. Responses that return error codes
(e.g., failure to write to storage) happen rarely, thus are hard to cover during
testing. Interactions with external services may time out. The dependence
on time further creates testing coverage issues.
\item Failures are non-deterministic events that may happen at any time.
Failure-injection tools are often hard to setup and control.
\item The specification requires consistency between the PoolManager state
(including its in-memory state as well as state stored in \AzureStorage) and
\VMSS. Writing such an assertion can be very cumbersome with traditional
means because it
spans multiple services.
\item The specification is a liveness property that requires the pool
to eventually reach steady state. Executions that get stuck in a loop
without making progress are violations of this property and hard to
capture using plain assertions.
\end{enumerate}

\noindent
The $\psharp$ testing methodology helped to overcome these challenges.
To the best of our knowledge, this is the first time a formal specification
was continuously tested against production code during its development.

\subsection{Mocking External Dependencies}

As mentioned in \sectref{Overview}, using $\psharp$ requires
writing \textit{mocks} of external services as well as all
external sources of non-determinism.
The \ABS engineering team wrote mocks for \AzureStorage, \VMSS,
the \ABS scheduler, and a basic system timer (used for encoding timeouts).

\paragraph{Mock \AzureStorage}
Writing a mock for \AzureStorage was easy.
It consists of roughly $800$ lines of code.
The mock has no internal concurrency (i.e., it executes sequentially)
but makes non-deterministic choices to expose various
error modes. \figref{TablesMock} shows a simple illustration
of the mock with a \texttt{Write} operation.
The entire store is modeled as an in-memory dictionary (\texttt{store}).
The \texttt{Write} operation can, for instance, either succeed
and write to storage, or it may return one of several error codes.
It can return an error code even after writing to the store successfully:
a possibility that can indeed happen with the real \AzureStorage
service. The mock also implements the \textit{ETag} matching logic (\sectref{ExternalServices}).

\vspace{2mm}
\paragraph {Mock DeploymentManager}
The mock for DeploymentManager, which wraps \VMSS, is around $1,200$ LoC.
This service has to handle requests for creating, growing and deleting
deployments, as well as for deleting specific VMs.
The mock uses an in-memory dictionary to track the deployments
and the VM instance names.
The mock non-deterministically returns HTTP errors including timeouts.
Like the real \VMSS service, the mock supports multiple levels
of failure (e.g., at the operation or HTTP level).
One key requirement was to ensure that the mock respects
\textit{idempotency} when the real service guaranteed it.
For example, once the mock returns success for an operation, then it has
to return success if the same operation request is issued again.

\begin{figure}[t]
\begin{lstlisting}[language={[Sharp]C},basicstyle=\scriptsize]
class MockAzureStorage
{
  // Map: StoreName -> PartitionName -> RowKey -> Entity
  Dictionary<string, Dictionary<string, Dictionary<string, Entity>>> Store;

  Response Write(WriteContext context, Entity entity)
  {
    // Does entity size exceed maximum allowed?
    if (IsEntityTooLarge(entity, context.StoreName)) {
      // Raise an error, PoolManager should never do this
      PSharpRuntime.Assert(false);
    } else if (PSharpRuntime.Nondet()) {
      return ErrorCode.TIMEOUT;
    } else if (PSharpRuntime.Nondet()) {
      return ErrorCode.STORAGE_ERROR;
    } else if (!IsEtagMatched(context)) {
      return ErrorCode.ETAG_CHECK_FAILED;
    } else {
      // Perform the write
      Store[context.StoreName] [context.PartitionName] [context.Key] = entity;

      if (PSharpRuntime.Nondet()) {
        // Return failure even when the write is done
        return ErrorCode.STORAGE_ERROR;
      } else {
        return StatusCode.OK;
      }
    }
  }
}
\end{lstlisting}
\vspace{-2mm}
\caption{A code snippet of the \AzureStorage mock.}
\vspace{-2mm}
\label{Fi:TablesMock}
\end{figure}

\paragraph {Mock scheduler}
The mock scheduler is roughly $600$ LoC and mimics the \ABS scheduler.
The job of the scheduler is to schedule tasks onto VMs.
From the perspective of the PoolManager, the scheduler has to handle requests
for adding or removing VMs and getting VMs that need recovery.
The mock scheduler has internal data structures that track pools and VMs
(essentially, a \texttt{struct} with multiple fields).
The mock scheduler can non-deterministically return failures or
remove a subset of VMs for a remove-VM request.
The mock is required to update the corresponding \AzureStorage entries as it accepts
or removes VMs.
This helps expose possible race conditions in the PoolManager,
when a VM that is picked for recovery is also removed during
a shrink operation in the PoolManager.
(The ETag logic of \AzureStorage helps discover such races.)

\begin{figure*}
\centering
\includegraphics[width=0.9\linewidth]{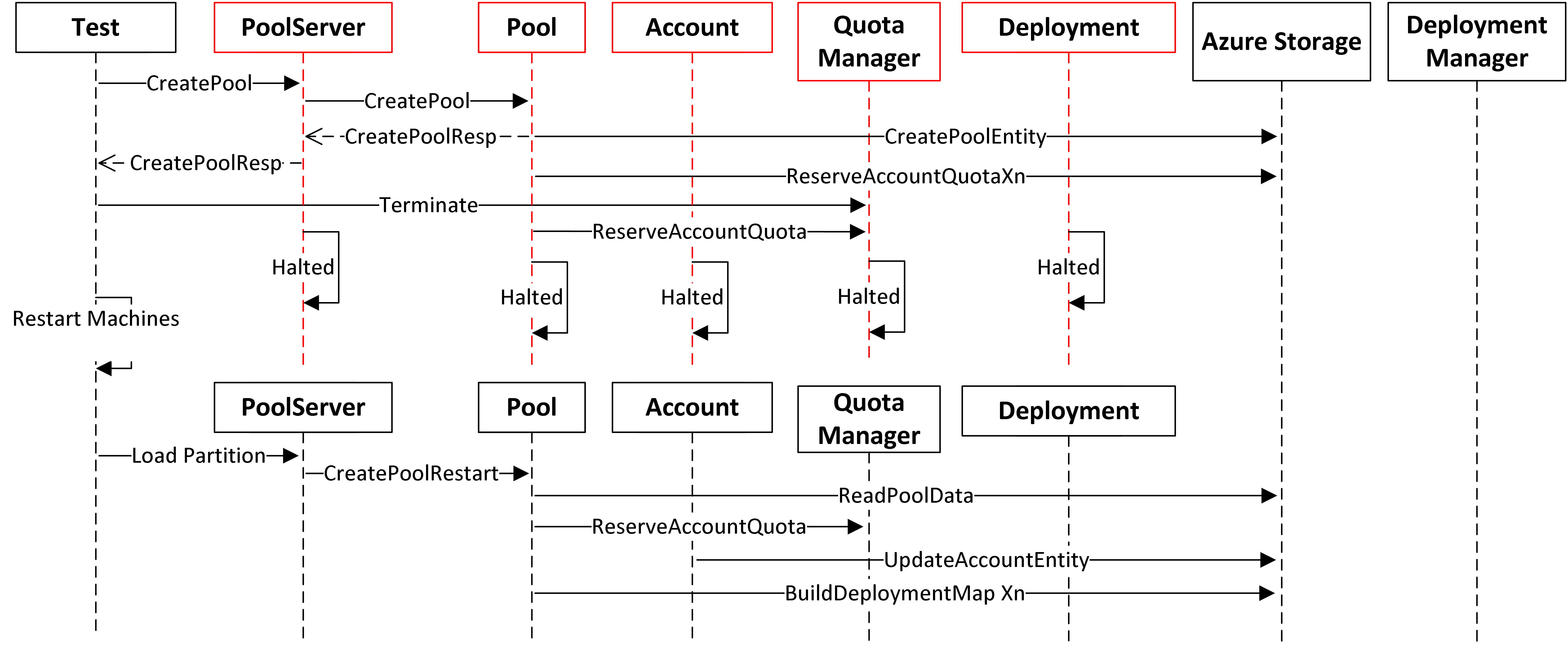}
\vspace{-3mm}
\caption{An illustration of failure injection into the PoolManager.}
\label{Fi:Failover}
\vspace{-1mm}
\end{figure*}

\paragraph{Mock timers}
All timers used in the PoolManager were also mocked.
\ABS engineers wrote an abstract \texttt{Timer} class.
During testing, \texttt{Timer} is implemented using a
$\psharp$ non-deterministic choice that can fire
the timeout at any point.
This helps abstract away time, expressing the fact that correctness
of the service does not rely on the particular timeout values chosen.
During production runs, an actual system timer is used for
implementing \texttt{Timer}.
Therefore, timeout values can be freely manipulated in the production
service in order to optimize performance.

As opposed to testing the PoolManager against production services
(e.g., \AzureStorage and \VMSS), creating mocks is additional
engineering cost that one must pay for using $\psharp$.
However, it also has several advantages.
First, it clearly lays down the assumptions that the
PoolManager is making of these external services. Any deviation from
reality can be captured and fixed in the mocks in order to avoid further regressions.
Second, all failure modes are made explicit in the mocks using non-deterministic
choices. This provides $\psharp$ tester with the hooks necessary to explore rare
or exceptional
behaviors. Third, the state stored in the external services is
captured in the (in-memory) mocks. Thus, asserting for
consistency between multiple distributed services is much easier.
Furthermore, the use of mocks allows testing to be fast: there is no need to wait
on wall-clock time in order to fire timers; no need to go over the
network to talk to external services, no need to write to disk to
survive (hard) failures, etc.

\ABS has various $\psharp$ tests that each exercise different APIs of the
PoolManager. All these tests work against the same \texttt{monitor} that
captures the specification of the system (\sectref{PoolManagerSpec}).
It is then up to the $\psharp$ tester to find a violation of the safety
and liveness properties specified by the monitor.
The $\psharp$ tester also checks for any assertions embedded
inside the code, as well as uncaught exceptions.

\subsection{Failover Testing Technique}
\label{Se:Failover}

The PoolManager failover logic is checked programmatically by mocking
failures themselves. The $\psharp$ tests and specifications do not change.
The \ABS team modeled failures by creating an \texttt{event} called
\texttt{Terminate} (from the perspective of $\psharp$, this is just a
user-defined \texttt{event} with no special meaning).
\figref{Failover} provides an illustration of the PoolManager execution
when a \texttt{Terminate} \texttt{event} is injected by the test harness
during a \texttt{CreatePool} operation.
When each \texttt{machine} in PoolManager receives a \texttt{Terminate}
\texttt{event} (at an arbitrary point), it forwards the \texttt{event}
to its \textit{children} \texttt{machines}
(forwarding of \texttt{Terminate} is not shown in the figure for simplicity),
waits for them to send a response and once all responses are received,
it halts itself.
This way, sending a \texttt{Terminate} \texttt{event} to the top-level
\texttt{machine} (\texttt{PoolServer}) ends up terminating
the entire PoolManager.
The \texttt{Terminate} \texttt{event} is only forwarded to the PoolManager
\texttt{machines} (in red), not the \texttt{machines} that mock external
services (because the test is for the PoolManager failover logic).
The failure injection, i.e., the action of sending a \texttt{Terminate}
\texttt{event}, is non-deterministic, thus the $\psharp$ tester will provide
coverage by exploring many different possibilities.

After all of the PoolManager \texttt{machines} halt, the test harness restarts
the service by re-creating the \texttt{PoolServer} \texttt{machine}.
When the \texttt{PoolServer} starts, it will read its state from storage,
where it will find the state from before the failure
(because the mock \AzureStorage \texttt{machine} survived the ``failure'').
Thus, the failover logic---\textit{the same logic written to handle real failures
in production}---kicks in and the PoolManager resumes
the \texttt{CreatePool} operation.

For most part, the relationship of a \texttt{machine} to its children
\texttt{machines} is obvious and follows the creation hierarchy:
if \texttt{machine} $A$ created \texttt{machine} $B$ then $B$ is $A$'s child.
In some cases, this is more involved, especially when \texttt{machines}
(legally) halt.
In this case, if a \texttt{machine}, say $A$, wishes to halt,
then it must first delegate the responsibility
of terminating its children \texttt{machines} to some other \texttt{machine}.
This was done using custom logic that the \ABS team designed
for the PoolManager. The same termination code is also used for legal
teardown of a PoolManager instance, so the team did
not see this effort as a test-only overhead.

A key advantage of this technique is that failovers are simply tested
at the level of program semantics. It does not require an actual setup
with hard failure injection that must crash and re-start the
process. The complete engineering of the tests simply
becomes a programming activity. It is then much easier for a developer
to control and observe failover coverage. For instance, a few lines of
code are enough to limit failover testing to one particular region of
code. There is no need of resorting to fuzzing or failure-injections
tools or a stress-test environment. Debugging is much easier as well because the
programmer is given a fully-replayable trace by the $\psharp$ tester, consisting
of the actions taken by the system both before and after the failure injection.

\section{Improvements in using \psharp}
\label{Se:ImprovementsToPSharp}

Supporting the development of production services required several engineering
enhancements to improve the programming, testing and debugging experience of
using $\psharp$.
% , which we contributed to the open-source $\psharp$ codebase.

\paragraph{Programming improvements}
The $\csharp$ language (which $\psharp$ extends) includes the
\texttt{async} and \texttt{await} keywords that make it easy to write
asynchronous code~\cite{asyncawait}.
Awaiting on an \texttt{async} call packages the current task as a continuation
and releases it from the executing thread, so that other tasks can be scheduled.
We enhanced $\psharp$ to allow \texttt{machine} handlers to be \texttt{async}:
this, in turn, allows handlers to call \texttt{async} APIs and \texttt{await} on their
result without blocking the underlying thread, enabling other \texttt{machines}
to be scheduled.

The $\psharp$ programming model discourages sharing objects between \texttt{machines}.
Message transfers are the only way for \texttt{machines} to synchronize,
which can be cumbersome for some tasks, compared to other forms of concurrency.
For example, consider the PoolManager task of maintaining the total rogue VM
count (\sectref{PoolManagerOperations}).
To maintain this count, the \texttt{QuotaManager} \texttt{machine} must
communicate with all \texttt{Deployment} \texttt{machines} in the system
and aggregate their individual counts.
Getting a count from each \texttt{machine} requires sending an \texttt{event}
to it, waiting for a response, and defining a handler for the response.
This not only increases the programming burden but is not efficient for
a simple task such as aggregating counts.

To remedy this situation, we developed a
library\footnote{\url{https://github.com/p-org/PSharp/tree/master/Source/SharedObjects}}
that allows a $\psharp$
\texttt{machine} to create a \textit{shared object} and freely pass its reference to
other \texttt{machines}.
These shared objects expose a
\textit{linearizable}~\cite{DBLP:journals/toplas/HerlihyW90} interface
so that multiple \texttt{machines} can issue operations on the
(same) shared object without concurrency issues.
In production, shared objects are implemented using an efficient
lock-free data structure.
When running under the $\psharp$ tester, they automatically resort to using
\texttt{machines} with message transfers so that the tester does not have to
understand any additional form of synchronization.
We implemented shared objects for common types such as counters and dictionaries.
The design of shared objects showcases the power of mocking:
the $\psharp$ programming model is not in conflict with low-level or
efficient concurrent programming of any form, it simply requires that any
concurrency outside of $\psharp$ be mocked for testing.

\paragraph{Testing improvements}
Even the smallest of $\psharp$ programs can have an astronomically large state space.
There are several \textit{search} strategies
developed in the research community that target finding common bug patterns fast.
Many of these strategies have complementary
strengths~\cite{DBLP:conf/sigsoft/DesaiQS15,DBLP:journals/topc/ThomsonDB16}.
The $\psharp$ tester includes multiple search strategies
and makes it easy to include new ones.
We enhanced the tester to run a portfolio of search strategies in parallel
so that engineers, who are likely unaware of these search algorithms,
do not have to worry about making the choice.

We further enhanced the $\psharp$ tester to parallelize it on the cloud.
We used \ABS itself: one can create a pool of VMs and run
the $\psharp$ tester in parallel on each VM.
$\psharp$ testing parallelizes easily: each instance of the tester simply
runs a different search strategy (with different parameters and seeds).
The typical requirement for many teams was to run the tester for
$10,000$ iterations, which could be easily achieved on a developer laptop
in a manner of minutes.
However, occasionally teams chose to run millions of iterations
for which cloud-scale testing was important.

\paragraph{Debugging improvements}
When the $\psharp$ tester finds a bug, it generates a trace file consisting
of all scheduling decisions as well as non-deterministic choices that it made.
This trace can be fed back to the tester, in which case,
it reproduces the same sequence of choices.
To improve the debugging experience, we enhanced the $\psharp$ tester by
allowing it to attach a debugger when replaying a trace.
Deterministically reproducing reported bugs in a concurrent and
non-deterministic system while being able to set breakpoints and
step-into the code, was a key value addition that has been appreciated
by all developers who have used $\psharp$ so far.

\section{Experience with using $\psharp$ in production}
\label{Se:Experience}

As discussed in \sectref{Introduction}, the positive experience of the \ABS team
using $\psharp$ invited attention from other teams in \Azure.
Besides PoolManager, there are eight other services built with $\psharp$ that
are currently live in production, with several more in the planning stage.
Many of these services share common characteristics with PoolManager:
asynchronously arriving requests that must be processed concurrently
in a non-blocking fashion, multiple distributed data sources that must be kept
consistent with each other, and interaction with several other services.

Each team echoed two \textit{key advantages} of using $\psharp$:
$(1)$ the \textit{actor programming model} (that $\psharp$ \texttt{machines} are based on)
allowed them to implement a service at a higher-level of abstraction, resulting in
a cleaner design and code that is easier to maintain, extend and explain to new team members;
and $(2)$ the \textit{high-coverage testing} allowed them to exercise many corner cases
and find several high-severity bugs before deployment.
The rest of this section summarizes the experience of all these teams from using
$\psharp$ to design, implement and test their cloud services.

\paragraph{Benefits in design and implementation}
Several teams reported that using the $\psharp$ actor programming model
helped them implement services that closely match their initial high-level
(whiteboard) design.
A senior \Azure engineer that used $\psharp$ gave us the following feedback:
\textit{``the design maps very closely to the actual code, usually I see
a much bigger delta between design and finished product''}.
Closing the gap between design and implementation, allowed teams to easily create
diagrams such as in \figref{CreatePool} that provide not only a detailed understanding
of the workflow implementing each operation, but also the expected communication
between $\psharp$ \texttt{machines}. These diagrams were useful in explaining the
design to other team members.
Other benefits of using the actor-based approach of $\psharp$ include:

\begin{itemize}
\item The \texttt{events} sent between \texttt{machines} have to be clearly defined in the
implementation. Further, no data can be shared between \texttt{machines} unless explicitly
sent via an \texttt{event}. Both of these helped improve code abstraction and readability.

\item $\psharp$ code is naturally non-blocking (asynchronous), so there is no need
for explicitly locking resources. Further, there is no need to manage a thread pool,
as this is done automatically by the $\psharp$ runtime. These benefits together were
a welcomed relief over typical multi-threaded code with threads and locks.

\item \texttt{Machines} are lightweight and event-based which can lead to
significant performance gains.
One of the teams reported that their previous design relied on polling, which consumed
too many CPU cycles. After rewriting their service to $\psharp$, the team was able to write
fully reactive code that allowed them to scale to much larger workloads. For instance,
they were able to hold up to $200,000$ $\psharp$ \texttt{machines}
before seeing the CPU reach 80\% utilization.
\end{itemize}

\noindent
It is worth noting that $\psharp$ was not perceived to be an ``arcane'' technology
only used and understood by the most senior engineers. A junior engineer that recently
joined one of the \Azure teams using $\psharp$ said:
\textit{``being a new developer to the team, one of the first few things I
worked on was $\psharp$, it was really quick to onboard and writing actual code
is simple and straightforward''}.

\paragraph{The importance of mocking}
Cloud services typically operate by communicating with their environment, which can consist
of other services, as well as resources such as network and storage.
To simplify the development process, teams would initially create \textit{interfaces}
for all external dependencies of their service, and then provide simple mock
implementations of these interfaces (e.g., the \ABS team created a mock for
\AzureStorage as seen in \figref{TablesMock}). Importantly, $\psharp$ mocks
allow developers to express nondeterministic behavior that can be controlled
during testing.
A large part of the development of each service was done against these mocks 
in a test environment. 

The efficacy of $\psharp$ testing relies on how closely mocks model the real behavior.
Any deviation can lead to missed bugs (when mocks do not exercise some possible behavior)
or even false alarms (when mocks exhibit some behavior that is not possible in reality).
Interestingly, each time there was an issue in production that was not found
by the $\psharp$ tester, it turned out to be either due to a missing test
(some workload was not exercised by the $\psharp$ tests) or an incomplete mock.
It was \textit{never} the case that the $\psharp$ tester could have found the bug,
but missed it because of lack of coverage.
In these cases, developers would add more tests or patch the mocks.
The teams knew about these tradeoffs before deciding to use $\psharp$.
Maintaining the mocks was an iterative process and deviations were fixed
over time as they got noticed. The initial mocks simply followed the available
online documentation and gradually got more detailed over time, and thus more
effective. This \textit{pay-as-you-go} model of writing mocks was important to avoid
front-loading the implementation with mocking effort.

No team reported mocking to be a burden, not only because they saw the value
that these mocks unlocked, but also because mocking is a common engineering exercise,
even without $\psharp$.
Some of the services that were mocked included:
\AzureServiceBus~\cite{AzureServiceBus},
\AzureCosmosDB~\cite{AzureCosmos},
\AzureStorage~\cite{azure-storage},
various networking services~\cite{AzureNetworking} and
resource providers~\cite{AzureResourceProviders}.
There was sharing of mocks between teams, but each team ended up owning its
own mocks so they could customize them in ways most relevant for their service.

To illustrate one example, the \Azure blockchain team built a
service~\cite{azureblockchain} using $\psharp$ that is
designed to hide the complexity of blockchains from users. It deals with issues
of submitting transactions exactly once, hiding forks and rollbacks from users,
etc. In order to test this service, the blockchain team wrote a mock of a
blockchain network itself that nondeterministically created forks and
rollbacks. The ability of authoring $\psharp$ tests for exercising such scenarios was very
valuable to the team.

\paragraph{The value of systematic testing}
Teams typically focused on only writing end-to-end $\psharp$ tests for their
services, as opposed to writing unit-tests for individual $\psharp$ \texttt{machines}.
This was enabled by the fact that systematic testing can deal with concurrency
and all declared sources of non-determinism. The testing process involves the $\psharp$
tester executing a test repeatedly from start to completion (what we call an iteration),
each time exploring execution paths using some specified
strategy~\cite{DBLP:conf/pldi/DeligiannisDKLT15}.
Without this support, engineers would need to write many more small unit-tests that exercise
individual components of their code along with custom assertions for each test.

A common specification among services was to validate liveness properties: each service
asserted that it would \textit{eventually} accomplish the client request, even in the
presence of failures.
$\psharp$ enabled writing such end-to-end specifications just once, and reuse them for
all relevant tests. The tests themselves only vary in the client workload they execute.
Teams reported that writing an end-to-end specification was much more concise than having
multitude of small tests. 
Further, $\psharp$ tests would typically run much faster than stress testing
or complex simulations. For example, exercising failover in PoolManager
(\sectref{Failover}) takes approximately two minutes for $10,000$ iterations
in an Intel Core i7 laptop with 4 cores and 16GB RAM.

Developers frequently executed $\psharp$ tests to validate safety
and liveness specifications as they made code changes, ensuring that the implementation
never regressed.
Some teams reported that the $\psharp$ tester helped them find several high-severity
bugs before deployment that would have been hard to find using conventional means,
and would have resulted in loss of business if they occurred in production.
In the words of a service architect, using $\psharp$ they
\textit{``found several issues early in the dev process, this sort of issues that would
usually bleed through into production and become very expensive to fix later''}.

Once the service code was reasonably functional in a test environment, a team
would re-implemented the mocked interfaces (as discussed above) to communicate with the actual external components.
Then the code---\textit{the same code that was systematically tested}---was deployed by
using this implementation of the interfaces.
As discussed in \sectref{Introduction}, teams reported that once a feature was tested
with $\psharp$, it would \textit{just work} when put into production.

\section{Related Work}
\label{Se:RelatedWork}

\paragraph{Systematic testing tools}
The research community has been long interested in finding bugs in
distributed systems.
Previous work showcased how \textit{systematic testing} (ST) tools
and search techniques can successfully find deep concurrency
bugs~\cite{musuvathi2008fair,musuvathi2008finding,yang2009modist,
simsa2011dbug,leesatapornwongsa2014samc}.
Many of these search techniques have been adapted almost directly
by $\psharp$~\cite{DBLP:conf/fast/DeligiannisMTCD16}.
However, it is critical to consider the nature in which these
techniques are exposed to a user.

In order to reduce user effort, ST tools have mostly targeted existing systems
without modification. 
This approach requires ST tools to take over \textit{all} sources of
non-determinism in these systems.
Obtaining such level of control is difficult because the API
surface of such systems can be very broad.
For example, \textsc{Chess}~\cite{musuvathi2008fair} targeted the testing
of concurrent multi-threaded programs on Windows.
To control thread interleaving, \textsc{Chess} had to interpose
at the Win32 API level (via stubs)
and reliably identifying all sources of thread synchronization.
It was necessary to get these stubs right, without which the tool would be
flaky or even deadlock, leading to user frustration.
The effort required to maintain these stubs was too large, and
\textsc{Chess} went out of support without seeing user adoption,
even though it found numerous tough bugs.
Instead of targeting unmodified systems, $\psharp$ spells out how
a new system must be built from the outset.
$\psharp$ concurrency is simple to control: its only about
\texttt{machine} creation and message passing.
Any use of external nondeterminism must be mocked: an exercise that
is much easier for a programmer who controls the design of their code.

Systems-level imposition can also be slow. For example,
\textsc{SAMC}~\cite{leesatapornwongsa2014samc} takes roughly $6$
hours for doing $1,000$ test iterations of Cassandra~\cite{Cassandra}
because it must bring up the actual database in each iteration and
inject failures via actual system crashes.
The use of mocks to model real-world interactions and failures
offers speed: roughly two minutes for $10,000$ iterations of
PoolManager with $\psharp$ (\sectref{Experience}).

\paragraph{Modeling languages}
Researchers have argued for principled design of distributed systems
through the use of modeling languages such as TLA+~\cite{lamport1994temporal}
or Promela~\cite{Holzmann:2011:SMC:2029108}.
TLA+ has been extensively used to model and specify distributed protocols
and algorithms~\cite{Newcombe:2015:AWS:2749359.2699417}.
One can apply inductive reasoning to these model (which is harder)
or use push-button model checkers (which is easy).
Modeling languages are useful for validating the high-level design of a system.
However, they do not help with the actual implementation.
As new features are added, the implementation often diverges from the initial
design that was modeled.
$\psharp$ bridges the gap between design and implementation, and what you test
is what you execute (\sectref{Experience}). 

\paragraph{Formally verified systems}
Recent research efforts have focused on developing
formally-verified systems~\cite{wilcox2015verdi,hawblitzel2015ironfleet}.
Building such systems involves using a high-level language
that can generate executable code, as well as contain logical assertions
that mark \textit{inductive} system invariants.
The inductiveness checks, as well as the check that
the invariants imply the system specification, are all discharged by a
theorem prover to establish the proof of correctness. Examples of such
systems include: crash-tolerant file systems~\cite{Chen:2017:VHC:3132747.3132776},
simple operating systems~\cite{hawblitzel2014ironclad},
distributed key-value stores~\cite{hawblitzel2015ironfleet}
and protocols~\cite{wilcox2015verdi,DBLP:conf/snapl/BhargavanBDFHHI17}.
Although this line of research is exciting, all of these systems have
been developed in academic settings. Inductive reasoning requires
deep understanding of formal logic and that is outside the scope of
education that most software developers receive. This constitutes the biggest
bottleneck for adoption of these practices in the industry.

$\psharp$ removes the need for theorem proving.
Developers must write specifications of their program to find bugs, but there is
no need for inductive reasoning. 
At the most, one must learn the concept of liveness properties, using the notion of
\textit{hot monitor states} (\sectref{PSharp}).
The emphasis of $\psharp$ in programmability allows engineering teams
to use $\psharp$ effectively without having a researcher in the loop.
This does imply that $\psharp$ guarantees are not as strong as full verification:
$\psharp$ testing is still an argument of coverage.
However, as this paper shows, $\psharp$ testing has (so far) not missed
a bug due to lack of coverage (\sectref{Experience}).

\paragraph{Previous work on P and $\psharp$}
Previous work on $\psharp$ focused on defining the
framework~\cite{DBLP:conf/pldi/DeligiannisDKLT15} and 
showcasing its bug-finding abilities~\cite{DBLP:conf/fast/DeligiannisMTCD16,
DBLP:conf/fmcad/MudduluruDDLQ17}, but only targeted existing systems.
In our experience, convincing teams to use $\psharp$ with just bug-finding
capabilities alone was not enough.
It was much easier to convince them once we had a success story (with \AzureBatch)
that demonstrated overall faster development time, along with increased service quality. 
It proved that the framework is mature, easy to learn and use.

The \texttt{P} language~\cite{DBLP:conf/pldi/DesaiGJQRZ13} is a different
design point compared to $\psharp$.
It consists of its own language and compiler, which increases
its barrier for adoption.
Although P has been successful for device driver development, integrating it with an underlying
infrastructure was very challenging.
With lack of libraries, IDE support, etc., it was hard convincing
engineering teams to take a dependence on it.
Our focus on cloud services (as opposed to \textit{all} asynchronous software)
also helped amplify our message within \Azure.

\bibliographystyle{unsrt}  
\bibliography{references}

\begin{thebibliography}{10}

\bibitem{forbes2018}
Forbes.
\newblock {83\% of enterprise workloads will be in the Cloud by 2020}.
\newblock
  \url{https://www.forbes.com/sites/louiscolumbus/2018/01/07/83-of-enterprise-workloads-will-be-in-the-cloud-by-2020},
  2018.

\bibitem{forbes2019}
Forbes.
\newblock {Public Cloud soaring to \$331{B} by 2022 according to Gartner}.
\newblock
  \url{https://www.forbes.com/sites/louiscolumbus/2019/04/07/public-cloud-soaring-to-331b-by-2022-according-to-gartner},
  2019.

\bibitem{amazon2012aws}
Amazon.
\newblock {Summary of the AWS service event in the US East Region}.
\newblock \url{http://aws.amazon.com/message/67457/}, 2012.

\bibitem{google2014outage}
Ben Treynor.
\newblock {GoogleBlog -- Today's outage for several Google services}.
\newblock
  \url{http://googleblog.blogspot.com/2014/01/todays-outage-for-several-google.html},
  2014.

\bibitem{tassey2002economic}
Gregory Tassey.
\newblock The economic impacts of inadequate infrastructure for software
  testing.
\newblock {\em National Institute of Standards and Technology, Planning Report
  02-3}, 2002.

\bibitem{musuvathi2008fair}
Madanlal Musuvathi and Shaz Qadeer.
\newblock Fair stateless model checking.
\newblock In {\em {Proceedings of the 29th ACM SIGPLAN Conference on
  Programming Language Design and Implementation}}, pages 362--371. ACM, 2008.

\bibitem{musuvathi2008finding}
Madanlal Musuvathi, Shaz Qadeer, Thomas Ball, Gerard Basler,
  Piramanayagam~Arumuga Nainar, and Iulian Neamtiu.
\newblock Finding and reproducing {Heisenbugs} in concurrent programs.
\newblock In {\em Proceedings of the 8th USENIX Conference on Operating Systems
  Design and Implementation}, pages 267--280. USENIX, 2008.

\bibitem{yang2009modist}
Junfeng Yang, Tisheng Chen, Ming Wu, Zhilei Xu, Xuezheng Liu, Haoxiang Lin, Mao
  Yang, Fan Long, Lintao Zhang, and Lidong Zhou.
\newblock {MODIST}: Transparent model checking of unmodified distributed
  systems.
\newblock In {\em {Proceedings of the 6th USENIX Symposium on Networked Systems
  Design and Implementation}}, pages 213--228. USENIX, 2009.

\bibitem{simsa2011dbug}
Ji\v{r}\'{\i} \v{S}im\v{s}a, Randy Bryant, and Garth Gibson.
\newblock {dBug}: Systematic testing of unmodified distributed and
  multi-threaded systems.
\newblock In {\em {Proceedings of the 18th International SPIN Conference on
  Model Checking Software}}, pages 188--193. Springer-Verlag, 2011.

\bibitem{leesatapornwongsa2014samc}
Tanakorn Leesatapornwongsa, Mingzhe Hao, Pallavi Joshi, Jeffrey~F. Lukman, and
  Haryadi~S. Gunawi.
\newblock {SAMC}: Semantic-aware model checking for fast discovery of deep bugs
  in cloud systems.
\newblock In {\em Proceedings of the 11th USENIX Conference on Operating
  Systems Design and Implementation}, pages 399--414. USENIX, 2014.

\bibitem{DBLP:conf/pldi/DeligiannisDKLT15}
Pantazis Deligiannis, Alastair~F. Donaldson, Jeroen Ketema, Akash Lal, and Paul
  Thomson.
\newblock Asynchronous programming, analysis and testing with state machines.
\newblock In {\em Proceedings of the 36th {ACM} {SIGPLAN} Conference on
  Programming Language Design and Implementation, Portland, OR, USA, June
  15-17, 2015}, pages 154--164, 2015.

\bibitem{psharpgithub}
The~P\# Team.
\newblock {P\#: A framework for rapid development of reliable asynchronous
  software}.
\newblock \url{https://github.com/p-org/PSharp}, 2019.

\bibitem{Agha:1986:AMC:7929}
Gul Agha.
\newblock {\em Actors: A Model of Concurrent Computation in Distributed
  Systems}.
\newblock MIT Press, Cambridge, MA, USA, 1986.

\bibitem{azure-batch}
Microsoft.
\newblock {Azure Batch: Cloud-scale job scheduling and compute management}.
\newblock \url{https://azure.microsoft.com/en-in/services/batch/}, 2019.

\bibitem{conf/cloud/VavilapalliMDAKEGLSSSCORRB13}
Vinod~Kumar Vavilapalli, Arun~C. Murthy, Chris Douglas, Sharad Agarwal, Mahadev
  Konar, Robert Evans, Thomas Graves, Jason Lowe, Hitesh Shah, Siddharth Seth,
  Bikas Saha, Carlo Curino, Owen O'Malley, Sanjay Radia, Benjamin Reed, and
  Eric Baldeschwieler.
\newblock Apache hadoop {YARN:} yet another resource negotiator.
\newblock In {\em {ACM} Symposium on Cloud Computing, {SOCC} '13, Santa Clara,
  CA, USA, October 1-3, 2013}, pages 5:1--5:16, 2013.

\bibitem{Hindman:2011:MPF:1972457.1972488}
Benjamin Hindman, Andy Konwinski, Matei Zaharia, Ali Ghodsi, Anthony~D. Joseph,
  Randy Katz, Scott Shenker, and Ion Stoica.
\newblock Mesos: A platform for fine-grained resource sharing in the data
  center.
\newblock In {\em Proceedings of the 8th USENIX Conference on Networked Systems
  Design and Implementation}, NSDI'11, pages 295--308, 2011.

\bibitem{azure-storage}
Microsoft.
\newblock {Azure Storage}.
\newblock \url{https://azure.microsoft.com/en-us/services/storage/}, 2019.

\bibitem{AzureVMSS}
Microsoft.
\newblock {Azure Virtual Machine Scale Sets}.
\newblock
  \url{https://azure.microsoft.com/en-in/services/virtual-machine-scale-sets/},
  2019.

\bibitem{azure-subs}
Microsoft.
\newblock {Azure Subscriptions}.
\newblock
  \url{https://docs.microsoft.com/en-us/azure/azure-subscription-service-limits},
  2019.

\bibitem{DBLP:conf/asplos/BurckhardtKMN10}
Sebastian Burckhardt, Pravesh Kothari, Madanlal Musuvathi, and Santosh
  Nagarakatte.
\newblock A randomized scheduler with probabilistic guarantees of finding bugs.
\newblock In {\em Proceedings of the 15th International Conference on
  Architectural Support for Programming Languages and Operating Systems,
  {ASPLOS} 2010, Pittsburgh, Pennsylvania, USA, March 13-17, 2010}, pages
  167--178, 2010.

\bibitem{DBLP:conf/fmcad/MudduluruDDLQ17}
Rashmi Mudduluru, Pantazis Deligiannis, Ankush Desai, Akash Lal, and Shaz
  Qadeer.
\newblock Lasso detection using partial-state caching.
\newblock In {\em 2017 Formal Methods in Computer Aided Design, {FMCAD} 2017,
  Vienna, Austria, October 2-6, 2017}, pages 84--91, 2017.

\bibitem{DBLP:conf/sigsoft/DesaiQS15}
Ankush Desai, Shaz Qadeer, and Sanjit~A. Seshia.
\newblock Systematic testing of asynchronous reactive systems.
\newblock In {\em Proceedings of the 2015 10th Joint Meeting on Foundations of
  Software Engineering, {ESEC/FSE} 2015, Bergamo, Italy, August 30 - September
  4, 2015}, pages 73--83, 2015.

\bibitem{DBLP:conf/fast/DeligiannisMTCD16}
Pantazis Deligiannis, Matt McCutchen, Paul Thomson, Shuo Chen, Alastair~F.
  Donaldson, John Erickson, Cheng Huang, Akash Lal, Rashmi Mudduluru, Shaz
  Qadeer, and Wolfram Schulte.
\newblock Uncovering bugs in distributed storage systems during testing (not in
  production!).
\newblock In {\em 14th {USENIX} Conference on File and Storage Technologies,
  {FAST} 2016, Santa Clara, CA, USA, February 22-25, 2016.}, pages 249--262,
  2016.

\bibitem{lamport1994temporal}
Leslie Lamport.
\newblock The temporal logic of actions.
\newblock {\em {ACM Transactions on Programming Languages and Systems}},
  16(3):872--923, 1994.

\bibitem{asyncawait}
Microsoft.
\newblock {Asynchronous programming with async and await in C\#}.
\newblock
  \url{https://docs.microsoft.com/en-us/dotnet/csharp/programming-guide/concepts/async/},
  2019.

\bibitem{DBLP:journals/toplas/HerlihyW90}
Maurice Herlihy and Jeannette~M. Wing.
\newblock Linearizability: {A} correctness condition for concurrent objects.
\newblock {\em {ACM} Trans. Program. Lang. Syst.}, 12(3):463--492, 1990.

\bibitem{DBLP:journals/topc/ThomsonDB16}
Paul Thomson, Alastair~F. Donaldson, and Adam Betts.
\newblock Concurrency testing using controlled schedulers: An empirical study.
\newblock {\em {TOPC}}, 2(4):23:1--23:37, 2016.

\bibitem{AzureServiceBus}
Microsoft.
\newblock {Azure Service Bus}.
\newblock \url{https://docs.microsoft.com/en-us/azure/service-bus-messaging/},
  2019.

\bibitem{AzureCosmos}
Microsoft.
\newblock {Azure Cosmos DB}.
\newblock \url{https://azure.microsoft.com/en-in/services/cosmos-db/}, 2019.

\bibitem{AzureNetworking}
Microsoft.
\newblock {Azure Virtual Network}.
\newblock \url{https://docs.microsoft.com/en-us/azure/virtual-network/}, 2019.

\bibitem{AzureResourceProviders}
Microsoft.
\newblock {Azure Resource Manager}.
\newblock \url{https://docs.microsoft.com/en-us/azure/azure-resource-manager/},
  2019.

\bibitem{azureblockchain}
Microsoft.
\newblock {Azure Blockchain Service}.
\newblock
  \url{https://docs.microsoft.com/en-us/azure/blockchain/service/overview},
  2019.

\bibitem{Cassandra}
Apache Foundation.
\newblock {Cassandra}.
\newblock \url{http://cassandra.apache.org/}, 2019.

\bibitem{Holzmann:2011:SMC:2029108}
Gerard Holzmann.
\newblock {\em The SPIN Model Checker: Primer and Reference Manual}.
\newblock Addison-Wesley Professional, 1st edition, 2011.

\bibitem{Newcombe:2015:AWS:2749359.2699417}
Chris Newcombe, Tim Rath, Fan Zhang, Bogdan Munteanu, Marc Brooker, and Michael
  Deardeuff.
\newblock How {Amazon Web Services} uses formal methods.
\newblock {\em Commun. ACM}, 58(4):66--73, March 2015.

\bibitem{wilcox2015verdi}
James~R. Wilcox, Doug Woos, Pavel Panchekha, Zachary Tatlock, Xi~Wang,
  Michael~D. Ernst, and Thomas Anderson.
\newblock Verdi: A framework for implementing and formally verifying
  distributed systems.
\newblock In {\em {Proceedings of the 36th ACM SIGPLAN Conference on
  Programming Language Design and Implementation}}, pages 357--368. ACM, 2015.

\bibitem{hawblitzel2015ironfleet}
Chris Hawblitzel, Jon Howell, Manos Kapritsos, Jacob~R Lorch, Bryan Parno,
  Michael~L Roberts, Srinath Setty, and Brian Zill.
\newblock {IronFleet}: Proving practical distributed systems correct.
\newblock In {\em Proceedings of the 25th Symposium on Operating Systems
  Principles}. ACM, 2015.

\bibitem{Chen:2017:VHC:3132747.3132776}
Haogang Chen, Tej Chajed, Alex Konradi, Stephanie Wang, Atalay \.{I}leri, Adam
  Chlipala, M.~Frans Kaashoek, and Nickolai Zeldovich.
\newblock Verifying a high-performance crash-safe file system using a tree
  specification.
\newblock In {\em Proceedings of the 26th Symposium on Operating Systems
  Principles}, SOSP '17, pages 270--286, New York, NY, USA, 2017. ACM.

\bibitem{hawblitzel2014ironclad}
Chris Hawblitzel, Jon Howell, Jay Lorch, Arjun Narayan, Bryan Parno, Danfeng
  Zhang, and Brian Zill.
\newblock Ironclad apps: End-to-end security via automated full-system
  verification.
\newblock In {\em USENIX Symposium on Operating Systems Design and
  Implementation (OSDI)}. USENIX - Advanced Computing Systems Association,
  October 2014.

\bibitem{DBLP:conf/snapl/BhargavanBDFHHI17}
Karthikeyan Bhargavan, Barry Bond, Antoine Delignat{-}Lavaud, C{\'{e}}dric
  Fournet, Chris Hawblitzel, Catalin Hritcu, Samin Ishtiaq, Markulf Kohlweiss,
  Rustan Leino, Jay~R. Lorch, Kenji Maillard, Jianyang Pan, Bryan Parno,
  Jonathan Protzenko, Tahina Ramananandro, Ashay Rane, Aseem Rastogi, Nikhil
  Swamy, Laure Thompson, Peng Wang, Santiago~Zanella B{\'{e}}guelin, and
  Jean~Karim Zinzindohoue.
\newblock Everest: Towards a verified, drop-in replacement of {HTTPS}.
\newblock In {\em 2nd Summit on Advances in Programming Languages, {SNAPL}
  2017, May 7-10, 2017, Asilomar, CA, {USA}}, pages 1:1--1:12, 2017.

\bibitem{DBLP:conf/pldi/DesaiGJQRZ13}
Ankush Desai, Vivek Gupta, Ethan~K. Jackson, Shaz Qadeer, Sriram~K. Rajamani,
  and Damien Zufferey.
\newblock {P:} safe asynchronous event-driven programming.
\newblock In {\em {ACM} {SIGPLAN} Conference on Programming Language Design and
  Implementation, {PLDI} '13, Seattle, WA, USA, June 16-19, 2013}, pages
  321--332, 2013.

\end{thebibliography}

\end{document}